\documentclass{sigplanconf}

\newcommand{\thetitle}{Variations on Variants}
\newcommand{\?}{\mathbin{\mathtt{?}}}

\usepackage{amsmath,amssymb,amsthm,array,centernot,enumerate,fancyvrb,infer,mathrsfs,mathwidth,relsize,stmaryrd,ulsy,xspace}
\usepackage[hyphens]{url}
\usepackage[usenames,dvipsnames]{xcolor}

\usepackage[utf8]{inputenc}
\usepackage[T1]{fontenc}
\usepackage{microtype}

\usepackage{zi4}
\DeclareMathAlphabet{\mathtt}{OT1}{zi4}{m}{n}
\makeatletter
\def\operator@font{\sf}
\makeatother

\usepackage{float}
\floatstyle{boxed}
\restylefloat{figure}

\usepackage{listings}
\lstnewenvironment{code}{}{}
\lstMakeShortInline{!}
\lstnewenvironment{codef}{\lstset{basicstyle=\ttfamily\small,numbers=left,xleftmargin=15px,numbersep=5pt}}{}

\theoremstyle{definition}

\newcommand{\seq}[1]{\vec #1}

\renewcommand{\eqref}[1]{(Equation~\ref{eqn:#1})}
\newcommand{\secref}[1]{(\S\ref{sec:#1})}
\newcommand{\figref}[1]{Figure~\ref{fig:#1}}

\newcommand{\ifne}[2]{\if\relax\detokenize{#1}\relax \else #2 \fi}

\newcommand{\Set}[1]{\{#1\}}

\newcommand{\entails}{\ensuremath{\then}}
\newcommand{\then}{\ensuremath{\Rightarrow}}
\newcommand{\If}[2]{\ensuremath{\ifne{#1}{#1 \then} #2}}

\newcommand{\predh}[2]{\ensuremath{#1\, #2}}
\newcommand{\predht}[2]{\ensuremath{\mathtt{#1}\, #2}}

\newcommand{\clause}[4]{\ensuremath{\ifne{#1}{#1 :} \ifne{#2}{\forall #2.\:} \If{#3}{#4}}}
\newcommand{\chain}[5]{\clause{#1}{#2}{#3}{#4} \; ; \; #5}

\newcommand{\gr}[1]{\lfloor #1 \rfloor}

{\[\begin{array}{lr@{\hspace{5px}}c@{\hspace{5px}}l}}
{\end{array}\]\ignorespacesafterend}

{\[\begin{array}{lrr@{\hspace{2px}}c@{\hspace{2px}}l}}
{\end{array}\]\ignorespacesafterend}

\newcommand{\todo}[1]{{\par\noindent\small\color{RoyalPurple} \framebox{\parbox{\dimexpr\linewidth-2\fboxsep-2\fboxrule}{\textbf{TODO:} #1}}}}

\newcommand{\isp}{\hspace{\infskip}}
\newcommand{\erule}[1]{(\textsc{#1})}

\usepackage[normalem]{ulem}
\usepackage{flushend}

\title{\thetitle}
\authorinfo{J. Garrett Morris}
           {The University of Edinburgh, UK}
           {Garrett.Morris@ed.ac.uk}

\begin{document}
\toappear{}

\maketitle

\begin{abstract}
  Extensible variants improve the modularity and expressiveness of programming languages: they allow
  program functionality to be decomposed into independent blocks, and allow seamless extension of
  existing code with both new cases of existing data types and new operations over those data types.

  This paper considers three approaches to providing extensible variants in Haskell.  Row typing is
  a long understood mechanism for typing extensible records and variants, but its adoption would
  require extension of Haskell's core type system.  Alternatively, we might hope to encode
  extensible variants in terms of existing mechanisms, such as type classes.  We describe an
  encoding of extensible variants using instance chains, a proposed extension of the class system.
  Unlike many previous encodings of extensible variants, ours does not require the definition of a
  new type class for each function that consumes variants.  Finally, we translate our encoding to
  use closed type families, an existing feature of GHC.  Doing so demonstrates the interpretation of
  instances chains and functional dependencies in closed type families.

  One concern with encodings like ours is how completely they match the encoded system.  We compare
  the expressiveness of our encodings with each other and with systems based on row types.  We find
  that, while equivalent terms are typable in each system, both encodings require explicit type
  annotations to resolve ambiguities in typing not present in row type systems, and the type family
  implementation retains more constraints in principal types than does the instance chain
  implementation.  We propose a general mechanism to guide the instantiation of ambiguous type
  variables, show that it eliminates the need for type annotations in our encodings, and discuss
  conditions under which it preserves coherence.
\end{abstract}

\category{D.3.3}
         {Programming Languages}
         {Language Constructs and Features}
         [Abstract data types]

\keywords extensible variants; row types; expression problem

\section{Introduction}

Modularity is a central problem in programming language design, and good modularity support has many
benefits.  Good modularity support improves extensibility and code reuse, saving programmer effort
and reducing the likelihood of bugs or infelicities in reimplemented functionality.  It also
provides for separation of concerns, assuring that conceptually independent features are implemented
independently, and simplifying refactoring of larger programs.

This paper studies extensible variants, a language mechanism that supports modular programming.
Extensible variants permit piecewise extension of algebraic data types with new cases, and support
code reuse in constructing and deconstructing values of extended data types. We present two
encodings of extensible variants, providing the same interface but using different extensions of the
Haskell class system (instance chains and closed type families).  Our goals in doing so are twofold.
First, we evaluate their expressiveness, by comparing them with row typing, a canonical approach to
extensible variants in functional languages.  Second, we use them as test cases to compare the
language mechanisms used in their definition.  We find that we can implement the same functions in
each encoding, and these functions are sufficient to express anything expressible with row types.
However, our encodings introduce a need for explicit type annotations (or type signatures) in
encoded terms where such annotations would not be necessary with row types.  We sketch a mechanism
that would eliminate the need for these type annotations.  Finally, while our encoding using closed
type families is as expressive as that using instance chains, a straightforward improvement of the
latter escapes easy translation to the former.

\paragraph{The expression problem.}

Wadler~\cite{Wadler98} proposed the expression problem as a benchmark for language expressiveness
and modularity.  The starting point is the definition by cases of a data type for arithmetic
expressions, and an operation over that data type.  For example, the data type might contain simple
arithmetic expression, and the operation might be evaluation.  The challenge is to extend the data
type with new cases and new operations, reusing the original code (without modification), and
preserving static type safety.

This framing of the expression problem may seem artificial.  However, similar problems arise
regularly in domains such as compilation.  For example, in implementing a Haskell compiler, we might
want to desugar surface language constructs, like special syntax for tuples, into a core syntax with
uniform notation for constructors.  The type of such a pass much capture the effect of the pass
(removing tuple syntax) and its requirements (the core syntax), but should not otherwise fix the
AST.  The encodings we present allow such typing; concretely, the pass would have the type
\[
  (\mathtt{Core} \olessthan (e \ominus \mathtt{Tuple})) \then \mathtt{Fix}\,e \to \mathtt{Fix}\,(e \ominus \mathtt{Tuple})
\]
where the $\olessthan$ constraint requires that the result type include the \texttt{Core} cases, and
the type operator $\ominus$ denotes removing cases from a type.

\paragraph{Implementing variants.}

Though definition of types by cases is standard in both functional and object-oriented languages,
the expression problem is challenging in either paradigm.  In many functional languages, adding new
cases to an existing data type requires changing the definition of the data type, and thus the
functions that use it.  In many object-oriented languages, adding new operations requires changing
the definition of the base class, and thus its subclasses.

There are at least two approaches to solving the expression problem in functional languages.  The
first approach, row typing~\cite{Wand87,Remy89,Remy92,Gaster96}, relies on an extension to the type
system specific to representing extensible records and variants.  The second approach represents
variants using generic binary coproduct and fixed point type constructors, and relies on overloading
to generalize injection and branching operations from the binary to the general
case~\cite{Swierstra08,Bahr14}.  This paper develops a new encoding of extensible variants, based on
the latter approach.  Our approach differs from previous encodings in several ways.  We permit the
use of arbitrarily structured coproducts in both introduction and elimination of extensible
variants, lifting technical restrictions present in many previous encodings.  More significantly, we
introduce a overloaded branching combinator, which can be seen as generalizing the categorical
notion of the unique arrow from a coproduct.  Unlike previous encodings, our approach does not
require that elimination of an extensible variants be defined using top-level constructs (like type
classes), and assures that elimination expressions cover all cases (unlike projection-based
approaches to variant elimination).  We give two implementations of our approach: one using instance
chains~\cite{MorrisJones10}, a proposed extension of the Haskell class system, and a somewhat more
verbose implementation using closed type families~\cite{Eisenberg14}, an existing feature of GHC.

\paragraph{Evaluating encodings.}

There is, of course, a cottage industry in encoding language features via increasingly cunning use
of type classes.  A critical question when evaluating any such encoding is how closely the encoding
matches the original language feature.  We examine how closely our encodings match approaches based
on row types.  While our system is sufficient to encoding arbitrary introduction and elimination of
extensible variants, losing no expressiveness compared to row-based systems, the same is not true of
the composition of introductions and eliminations.  We identify a typing ambiguity that appears in
all the encodings we know of, not just in ours, requiring the programmer to provide explicit type
annotations not required by row type systems.  Resolving this ambiguity requires the compiler to
make seemingly arbitrary choices of type instantiation during type checking; we propose a new
mechanism to guide this choice, and discuss the conditions under which the use of this mechanism
does not cause incoherence in the resulting programs.

\paragraph{Contributions.}

In summary, this paper contributes:
\begin{itemize}
\item A new approach to encoding extensible variants in Haskell, based on overloaded injection and
  branching combinators;
\item Implementations of this approach using instance chains and closed type families; and,
\item A comparison of these systems with each other and with row type systems, and a proposed
  language mechanism to address the expressiveness gap between them.
\end{itemize}
To that end, we begin by describing existing approaches to providing extensible variants in
functional languages, based on row types or overloaded injection functions~\secref{background}.  We
then describe our approach, and implement it using instance chains~\secref{ic}.  We show how our
approach can be used to solve the expression problem, and show how it can give precise types to
desugaring steps in programming language implementations.  We compare our approach to systems built
on row types~\secref{coherence}.  We conclude that all the existing approaches to encoding
extensible variants in Haskell suffer from typing ambiguities, requiring programmers to add type
annotations not required by row type systems, and propose a simple mechanism to eliminate the need
for such annotations.  We then translate our implementation of extensible variants to use closed
type families instead of instances chains~\secref{tf}.  This translation illustrates the
similarities and differences between the two language mechanisms.  We conclude by discussing
related~\secref{related} and future~\secref{conclusions} work.

\section{Rows and Variants}\label{sec:background}

\subsection{Row Typing and Qualified Types}

Wand~\cite{Wand87} introduced row types as a mechanism to type objects with inheritance.  In his
approach, the language of types is extended with rows, or sequences of labeled types
$\ell_1:\tau_1,\dots,\ell_n:\tau_n$.  Records and variants are constructed from rows; a record of
type $\Pi(\ell_1:\tau_1,\dots,\ell_n:\tau_n)$ has fields $\ell_1$ through $\ell_n$ with
corresponding types, while a variant of type $\Sigma(\ell_1:\tau_1,\dots,\ell_n:\tau_n)$ is given by
one of the labels $\ell_i$ and a value of type $\tau_i$.  Wand introduced row variables $\rho$ to
permit polymorphism in row-typed operations.  For example, the injection operator for a label $\ell$
would have the type $\alpha \to \Sigma(\rho[\ell \gets \alpha])$, where $\alpha$ ranges over types,
$\rho$ ranges over rows, and $\rho[\ell\gets\alpha]$ denotes the result of adding (or replacing)
label $\ell$ with type $\alpha$ to $\rho$.  Wand provides a branching combinator of type $(\alpha
\to \beta) \to \beta \to \Sigma(\rho[\ell \gets \alpha]) \to \beta,$ where the second argument is a
  default (or else) branch.

Wand's types do not track those labels not present in rows; thus, the type $\rho[\ell \gets \tau]$
may either add a new pair $\ell: \tau$ to $\rho$ or replace an existing pair $\ell: \tau'$.  As a
consequence, some programs in his calculus do not have principal types.  R\'emy~\cite{Remy89,Remy92}
proposed a variant of Wand's system that associates labels with flags rather than with types
directly; each flag $\phi$ is either $pre(\tau)$, indicating that the label is present with type
$\tau$, or $abs$, indicating that the label is absent.  For example, in R\'emy's calculus the
injection function for label $\ell$ has type $\alpha \to \Sigma(\ell:pre(\alpha);\rho)$, indicating
that label $\ell$ must be present in the result type, and the branching combinator for label $\ell$,
$case_\ell$, is given the type
\[
  (\alpha \to \gamma) \to (\Sigma(\ell: abs; \rho) \to \gamma) \to \Sigma(\ell: pre(\alpha); \rho) \to \gamma,
\]
where in each case $\ell: \phi; \rho$ denotes the extension of row $\rho$ with the pair $\ell:
\phi$, and is defined only if $\rho$ does not already contain some type labeled by $\ell$.  Note the
refinement compared to how branching is typed in Wand's calculus: in the expression $case_{\ell} \,
M \, N \, P$ we can assume that option $\ell$ is not present in the argument to $N$.

Gaster and Jones~\cite{Gaster96} propose a variant of row typing that represents negative
information using predicates on (row) types.  As a consequence, their system captures the
expressiveness of R\'emy's system but can use a simpler form of row types.  For example, the
injection operator in their system has type
\[
  (\rho \setminus \ell) \then \alpha \to \Sigma(\ell:\alpha; \rho)
\]
and their branching operator has type
\[
  (\rho \setminus \ell) \then (\alpha \to \gamma) \to (\Sigma(\rho) \to \gamma) \to \Sigma(\ell:\alpha; \rho) \to \gamma,
\]
where in each case the constraint $\rho \setminus \ell$ is satisfiable only if $\rho$ does not
contain a label $\ell$.  Unlike R\'emy's approach, the system of Gaster and Jones does not need
flags, and does not impose non-duplication constraints on the formation of rows.  As it builds on
Jones's system of qualified types~\cite{Jones92}, Gaster and Jones's system enjoys principal types,
type inference, and easy integration with type classes and other features expressible with qualified
types.  Two properties of their type system are central to their principality and type inference
results.  First, like other row type systems, they consider types equivalent up to rearrangement of
rows.  Second, they show that, in addition to most general unifiers, they can compute most general
inserters, or the most general substitutions for row variables than guarantee the inclusion of
particular labeled types.

\subsection{Modular Interpreters and Data Types \`a la Carte}

Wand originally introduced row types as a generalization of binary products and coproducts.  An
alternative approach to extensible variants is to use binary coproducts directly, but to generalize
the injection and branching operators.  Systems based on this approach differ from row-typing
approaches in two ways.  First, they tend not to rely on labeling types.  With the addition of
suitable type-level machinery for labels, however, they can be straightforwardly adapted to work on
labeled types.  Second, binary coproducts are not identified up to associativity and commutativity.
Thus, a central concern for these systems is not introducing distinctions among equivalent (but
rearranged) coproduct types.

Liang et al.~\cite{Liang95} gave an early example of this approach, as part of describing a modular
approach to building language interpreters.  They represent larger types as (right-nested)
coproducts of smaller types; for example, a term language including arithmetic (!TermA!) and
functional (!TermF!) terms would be described by !OR TermA (OR TermF ())! (where !OR! is their
coproduct type constructor).  They define a type class !SubType! to simplify working with
coproducts; $\predht{SubType}{\tau\,\upsilon}$ holds if $\upsilon$ is a right-nested coproduct and
$\tau$ is one of its left-hand sides; it provides methods $\mathtt{inj} :: \tau \to \upsilon$ to
inject values of component types into the coproduct type and $\mathtt{prj} :: \upsilon \to
\mathtt{Maybe}\,\tau$ to project values of component types from values of the coproduct type.  For
example, their system would provide functions
\begin{code}
inj :: TermA -> OR TermA (OR TermF ())
prj :: OR TermA (OR TermF ()) -> Maybe TermF
\end{code}
Liang et al.\ define type classes for operations on variant types, such as interpretation, with
instances for each term type and a generic instance for coproducts, wrapping the use of !prj!.
Their approach does not directly address extensible variants: recursion is hard-wired into the term
types.

Swierstra~\cite{Swierstra08} proposed another approach to extensible variants in Haskell, which he
called ``Data Types \`a la Carte''.  He defines variants by combining binary coproducts with Sheard
and Pasalic's~\cite{Sheard04} approach to open recursion (or ``two-level types'').  Consequently,
individual cases in his approach are functors, rather than ground types, in which the functor's
argument is used for recursive cases.  Similarly, rather than defining coproducts of ground types,
he defines coproducts of functors (written !f :+: g!).  Finally, he uses a fixed point constructor
!Fix! to construct types from functors. For example, in his system the types !TermA! (for arithmetic
expressions) and !TermF! (for functional expressions) would be functors, and the combined expression
type would be written !Fix (TermA :+: TermF)!.

Like Liang et al., Swierstra defines a class, called !(:<<:)!, to generalize injection into
(right-nested) coproducts.  His !(:<<:)! class defines an injection function but not a projection
function; he relies on type classes to implement functions that consume variants.  Thus, his system
provides functions like
\begin{code}
inj :: TermF e -> (TermA :+: TermF) e
\end{code}
Unlike the !SubType! class, !(:<<:)! is reflexive, and so can have
\begin{code}
inj :: TermF e -> TermF e
\end{code}
This avoids the need for ``terminator'' like !()! in the types of Liang et al.  As a consequence,
however, Swierstra's instances are ambiguous for predicates of the form !(f :+: g) :<<: h!.

Bahr~\cite{Bahr14} gives an extension of Swierstra's approach and an implementation using closed
type families.  He follows Liang et al. in giving a version of the subtype class that provides both
injection and projection operators; thus, his encoding does not require each elimination of
extensible variants to be defined by a new type class.  However, the projection-based approach does
not guarantee that pattern matches are complete.  Bahr finds an interesting solution to this
problem.  He defines his injection function with sufficient generality that he can use it to permute
the structure of coproducts, and defines a !split! operator that rearranges its argument to surface
desired cases.  By then using the standard Haskell \texttt{case} construct on the results of
!split!, Bahr can define extensible but complete branching.  His approach to defining !split!
introduces ambiguity, however, requiring the programmer to add explicit type signatures or proxy
arguments.

\section{Extensible Variants with Instance Chains}\label{sec:ic}

In this section, we describe another approach to encoding extensible variants.  We begin from the
same coproduct and fixed point constructors used by Swierstra~\cite{Swierstra08}.  However, our
approach differs from his in two important ways.  First, we define a more expressive inclusion class
!(:<:)!.  Our class is reflexive, permits both left-nesting and right-nesting in coproduct
construction, and excludes coproducts with repeated types (as an overloaded injection function into
such a coproduct must depend on an essentially arbitrary choice of which instance of the source type
to prefer).  Second, we define a generic expression-level branching combinator !(?)! instead of
defining new type classes for each deconstruction of an extensible variant type.

The implementations in this section rely on functional dependencies (introduced by
Jones~\cite{Jones00}) and instance chains (introduced by Morris and Jones~\cite{MorrisJones10}).
Functional dependencies capture dependency relationships among class parameters, directing the
instantiation of type variables appearing in class predicates.  Instance chains extend the Haskell
class systems with negated predicates and alternative instances, and base instance selection on the
provability of an instance's hypotheses rather than just the form of its conclusion.  We will
discuss the syntax, interpretation and motivation of these constructs as they are encountered in the
implementations; formal descriptions of instances chains are available in our previous
work~\cite{MorrisJones10,Morris13}.  Later~\secref{tf}, we will demonstrate how these
implementations can be translated to use closed type families~\cite{Eisenberg14}, a related feature
of the GHC type system.  This translation introduces not-insignificant complication, however,
motivating us to present both versions.

\subsection{Sums and Open Recursion}\label{sec:sums-open-recursion}

\begin{figure}
\begin{codef}
data Fix e     = In (e (Fix e))
data (f :+: g) e = Inl (f e) | Inr (g e)

(.?.) :: (f e -> a) -> (g e -> a) -> (f :+: g) e -> a
(f .?. g) (Inl x) = f x
(f .?. g) (Inr x) = g x
\end{codef}
\caption{Data types for variants and recursion.}\label{fig:types}
\end{figure}

Our first problem is to define the form of extensible variants.  We broadly follow the approach used
by Liang et al.~\cite{Liang95} and Swierstra, using a generic coproduct constructor to combine
individual type constructors and a fixed point combinator to implement recursive types.  The
definitions are given in \figref{types}.  For functors $\tau$ and $\tau'$, the coproduct type
$(\tau \oplus \tau') \upsilon$ has injectors !Inl! for $\tau \upsilon$ values and !Inr! for
$\tau' \upsilon$ values.  We also define a branching combinator !(.?.)! which, given two functions
of type $\tau \upsilon \to \upsilon'$ and $\tau' \upsilon \to \upsilon'$ respectively, produces a
function of type $(\tau\oplus\tau') \upsilon \to\upsilon'$.

\begin{figure}
\begin{codef}
data Const e   = Const Int
data Sum e     = Plus e e
data Product e = Times e e

type E1  = Fix (Const :+: Sum)
type E1' = Fix (Sum :+: Const)
type E2  = Fix ((Const :+: Sum) :+: Product)
\end{codef}
\caption{Expression constructors and expression types}\label{fig:expr}
\end{figure}

We will use a simple instance of the expression problem as a motivating example throughout this
section.  For this example, we will start with an expression language that contains only integer
constants and addition; we will demonstrate how we can add support for multiplication to this
language.  \figref{expr} gives the AST constructors for our language; note that the constant case
must be expressed as a functor, even though it contains no recursive instances of the expression
type.  We also give two types for terms in the initial form of the language (!E1! and !E1'!) and one
type for terms in the extended form (!E2!)

This example makes apparent the difficulties with using binary coproducts directly.  For example,
the form of a constant term differs in each version of the language, depending on the order of
summands in the coproduct used to define the term type:
\begin{code}
In (Inl (Const 1)) :: E1
In (Inr (Const 1)) :: E1'
In (Inl (Inl (Const 1))) :: E2
\end{code}
Clearly, code written for !E1! or !E1'! cannot be reused at type !E2!; similar problems would arise
in code that uses !(.?.)! to consume values of coproduct type.  In the remainder of the section, we
will implement type-directed versions of injection and branching combinators, allowing uniform
expression of terms of the various expression languages.

\subsection{Injection}\label{sec:inject}

We begin by describing our polymorphic injection function, which can be seen as a generalization of
the primitive injectors !Inl! and !Inr!.  Our goal is to implement something that looks like
Swiestra's injection function, but whose semantics are closer to the primitives of Gaster and
Jones~\cite{Gaster96}; that is, it should not impose particular structural requirements on
coproducts, and it should exclude coproducts with duplicate types, as its behavior in such cases is
essentially arbitrary.

\begin{figure}
\begin{codef}
class In f g

instance f `In` f
else f `In` (g :+: h) if f `In` g
else f `In` (g :+: h) if f `In` h
else f `In` g fails
\end{codef}
\caption{Membership test for sums.}\label{fig:in}
\end{figure}

Central to Gaster and Jones's approach are lacks constraints $\rho \setminus \ell$, denoting that
row $\rho$ does not contain a type labeled by $\ell$.  We must define a similar constraint; in our
setting, we find it easier to define a constraint that holds when a type is a component of a given
coproduct, and then use its negation to express the lacks constraint.

Our positive constraint is defined in \figref{in}.  We begin by introducing a two-parameter class
!In! (line 1); as we are capturing type-level structure, this class has no methods.  We populate the
class using an instance chain.%
\footnote{An instance chain is an ordered sequences of alternative instances, separated by
  \texttt{else}; later instances in the chain are used only if earlier instances do not apply.  Note
  that, in the syntax of instance chains, we write the conclusion before the hypotheses
  (\scantokens{!p if P!}\unskip) rather than after (\scantokens{!P => p!}\unskip); we find this
  makes instances easier to read as the list of hypotheses grows.}
The first instance (line 3) specifies that !In! is reflexive.  The remaining instances in the chain
will be used only when the first does not apply; that is, only when the two arguments to !In! do not
unify.  The second and third instances (lines 4-5) define when a type is a contained by a coproduct:
either because it is a component of the left-hand or right-hand summand.  The final instance (line
6) specifies that if none of the previous cases apply, the type !f! is not in !g!.%
\footnote{The latter three instances illustrate the other two aspects of instance chains.
\begin{itemize}
\item First, we introduce negated predicates to the class system: the last instance asserts the
  negation of \scantokens{!In f g!}\unskip.  Note that Haskell's module system necessitates an
  intuitionistic treatment of negation: simply because, for example, $\predht{Eq}{\tau}$ is not
  provable where a term is typed does not mean that term will not be used in a context where
  $\predht{Eq}{\tau}$ is provable.  The predicate \scantokens{!In f g fails!} does not simply assert
  that \scantokens{!In f g!} cannot be proven now, but that it will not become provable in any
  module that imports this one.
\item Second, we stated that later instances in a chain are tried only if earlier instances do not
  apply.  With instance chains, we consider that an instance does not apply to a predicate either if
  its conclusion does not match the predicate or if (at least) one of its hypotheses can be
  disproven.  For example, in attempting to prove the predicate \scantokens{!In B (A :+:
    B)!}\unskip, we would begin by trying the second instance; this would require proving that
  \scantokens{!In B A!}\unskip.  However, we can disprove this, using the fourth instance.  We would
  then apply the third instance to the original predicate, which shows that the predicate holds.
\end{itemize}}
As defined, this class seems very close to the one we wanted.  Unfortunately, it does not exclude
coproducts with repeated types (we can prove !In f (f :+: f)!) and we do not know of any
straightforward modification of it that does.  For example, one might hope to add an instance
\begin{code}
else f `In` (g :+: h) fails if f `In` g, f `In` h
\end{code}
between lines 3 and 4; however, while this excludes !In f (f :+: f)!, it does not exclude
!In f (f :+: (f :+: f))!.  Note that, because of the ordering of lines 4 and 5, constraints may not
be discharged as soon as one might hope.  For example, the constraint !A `In` (f :+: A)!, where type
variable !f! is otherwise unconstrained, cannot be discharged.  This is because the instance at line
4 matches, but its hypotheses can neither be proved or disproved.  However, the predicate will be
discharged as soon as !f! is instantiated.

\begin{figure}
\begin{codef}
class f :<: g where
  inj :: f e -> g e

instance f :<: f where
  inj = id
else f :<: (g :+: h) if f :<: g, f `In` h fails where
  inj = Inl . inj
else f :<: (g :+: h) if f :<: h, f `In` g fails where
  inj = Inr . inj
else f :<: g fails
\end{codef}
\caption{Overloaded injection function}\label{fig:inj}
\end{figure}

We can now define our inclusion class !(:<:)! and injection function !inj!, as shown in
\figref{inj}.  We begin by declaring the !(:<:)! class (lines 1--2); we include only an injection
method, as we will define branching separately.  We populate the class with another instance chain.
The first instance in the chain (lines 4-5) makes !(:<:)! reflexive; the injection function in this
case is trivial.  The next two instances handle (non-reflexive) injection into coproducts.  The
first case (lines 6--7) handles injection on the left-hand side of the coproduct (i.e., into !g!);
we insist that type !f!  not also appear on the right-hand side (i.e., in !h!), internalizing the
lacks constraint present in the typing of Gaster and Jones's primitives.  The injection function
from !f e! into !(g :+: h) e! is the injection from !f e! into !g e! followed by !Inl!, where we
rely on a recursive call to !inj! to determine the initial injection.  The second case (lines 7--8)
is parallel but for the right-hand side.  The final case rules out any other injections.  Note that
this final case is not strictly necessary---we never rely on proving !f :<: g fails!.  However, it
assures that the definitions of !In! and !:<:! remain synchronized.

We demonstrate the injection function by defining several terms in our simple expression languages.
We begin by defining a shorthand for injection into fixed points of functors:
\begin{code}
inj' = In . inj
\end{code}
We define a term that makes use of only constants and addition:
\begin{code}
x = inj' (inj' (Const 1) `Plus` inj' (Const 2))
\end{code}
Because of the overloading of !inj!, we can use !x! at any type that contains both !Const! and
!Plus!; that is, the principal type of !x! is
\begin{code}
(Const :<: f, Sum :<: f) => Fix f
\end{code}
Note that all the languages we defined (!E1!, !E1'!, and !E2!) satisfy these constraints.  Thus, we
can use !x! as a term in any of those languages, without having to change the definition of !x!.
For example, we can define a term using products, but including !x! as a subterm:
\begin{code}
y = inj' (inj' (Const 3) `Times` x)
\end{code}
The principal type of !y! includes the constraints required by !x!, but also requires !Product!; thus,
we can use !y! at type !E2! but not !E1! or !E1'!.  We could, however, use !y! at any permutation of
the constructors of !E2! or any larger type.

\subsection{Branching}\label{sec:branch}

\begin{figure}
\begin{codef}
class f :-: g = h where
  (?) :: (g e -> a) -> (h e -> a) -> f e -> a

instance (f :+: g) :-: f = g where
  m ? n = m .?. n
else (f :+: g) :-: g = f where
  m ? n = n .?. m
else (f :+: g) :-: h = (f :-: h) :+: g if h `In` g fails where
  m ? n = (m ? (n . Inl)) .?. (n . Inr)
else (f :+: g) :-: h = f :+: (g :-: h) if h `In` f fails where
  m ? n = (n . Inl) .?. (m ? (n . Inr))
\end{codef}
\caption{Overloaded branching combinator.}\label{fig:branching}
\end{figure}

The second part of the expression problem is to define extensible functions over the already-defined
(extensible) types.  While it is possible to do so using only existing features of Haskell, as
Swierstra does, this relies on implementing each operation that consumes variants as a type class
itself.  Instead, we define an overloaded branching combinator, generalizing the primitive branching
combinator !(.?.)!.  Our goal is the branching combinator of Gaster and Jones: !m ? n!  defines a
function on coproducts type where !m! describes its behavior on one summand of the coproduct and !n!
describes its behavior on the remainder of the coproduct.  This definition will have both type and
value level components.  At the type level, we must define what it means to remove one component of
a coproduct.  At the value level, we must define how the branching combinator combines !m! and !n!,
given that the case handled by !m! may be nested among those handled by !n!.

\figref{branching} gives our definition of the branching operator.  We begin by declaring class
!(:-:)! (lines 1--3).%
\footnote{We adopt several syntactic conventions for functional dependencies suggested by Jones and
  Diatchki~\cite{JonesDiatchki08}.  First, we write \scantokens{!f :-: g = h!} in the class
  declaration to denote that \scantokens{!:-:!} is a three-parameter class in which the first and
  second parameters determine the third (that is, there is a functional dependency \scantokens{!f g
    -> h!}\unskip).  Second, we will regularly write $\tau\ominus\tau'$ as a type; this denotes a new type
  variable $v$ such that the constraint $\predh{(\ominus)}{\tau\,\tau'\,v}$ holds.}
The predicate $\tau\ominus\tau'=\upsilon$ holds if $\tau$ is a coproduct containing summand $\tau'$
and $\upsilon$ describes the remaining summands of $\tau$ after removing $\tau'$.  For example, we
would expect that:
\begin{code}
(Int :+: Bool) :-: Bool = Int
((Int :+: Char) :+: Bool) :-: Char = Int :+: Bool
\end{code}
The branching operator !(?)! combines !m!, an operation on one summand !g e!, with !n!, an operation
on the remainder of the coproduct !h e!, to give an operation on the entire coproduct !f e!.  We
begin by considering the base cases.  Subtracting !f! from !f :+: g! leaves !g! (lines 4--5); in
this case, the overloaded branching operator is equivalent to the primitive branching operator.
Alternatively, subtracting !g! from !f :+: g! leaves !f! (lines 6--7); in this case, the branching
operator is the flip of the primitive branching operator.  The recursive cases are more interesting.
The left-recursive case (lines 8--9) describes the case when !h! is a component of the left-hand
summand !f!; in this case, the result of removing !h! from !f :+: g! is given by \scantokens{!(f :-:
  h) :+: g!}\unskip.  To avoid ambiguity, we insist that !h! not also appear in !g!; this also
simplifies the definitions of these cases.  To define the branching operator for this case, we
consider the possible input values (of type !f :+: g!).  One the one hand, the input value may be of
type !f!; in this case, it is either of type !h!, and is thus handled by !m!, or is of type
\scantokens{!f :-: h!}\unskip, and is handled by the left branch of !n! (i.e., by !n . Inl!).  Thus,
the behavior of !m ? n!  for arguments of type !f! is given by !m ? (n . Inl)!.  Alternatively, the
input may be of type !g!; in this case, it is handled by the right branch of !n! (i.e., by
\scantokens{!n . Inr!}\unskip).  These two cases are combined using the primitive branching operator
!(.?.)!.  The right-recursive case (lines 10--11) is parallel.

To demonstrate the !(?)! operator, we define several operations on our simple expression languages.
First, we consider evaluation.  We begin by defining evaluation functions for each case; in addition
to the term being evaluated, each function takes an additional argument !r! to handle recursive
expressions.
\begin{code}
evalConst   (Const x) r   = x
evalSum     (Plus x y) r  = r x + r y
evalProduct (Times x y) r = r x * r y
\end{code}
We define a helper function that unrolls the !Fix! data type:
\begin{code}
cases cs = f where f (In e) = cs e f
\end{code}
Finally, we can combine the functions for individual cases above to define evaluators.  For example,
the following function can be used to evaluate terms of either type !E1! or type !E1'!:
\begin{code}
eval1 = cases (evalConst ? evalSum)
\end{code}
The inferred type for !eval1! is
\begin{code}
(f :-: Const = Sum) => Fix f -> Int
\end{code}
Note that the order of cases is irrelevant; we could equally well have used !evalSum ? evalConst!.
To handle terms of type !E2!, we include the case to handle products:
\begin{code}
eval2 = cases (evalProduct ?
               (evalSum ? evalConst))
\end{code}
As we would hope, we are able to use the same code for each case, regardless of the order of cases
or the other cases appearing in the data type.

Instead of defining the entire evaluator at once, we might prefer to begin by desugaring complex
language constructs into simpler ones.  Suppose we had an additional term type for squares:
\begin{code}
data Square e = Square e
\end{code}
We can define a function that desugars !Square e! into !Times e e!:
\begin{code}
desugarSqr = cases (sqr ? def) where
  sqr (Square e) r = inj' (Times (r e) (r e))
  def e r = In (fmap desugarSqr e)
\end{code}
The default case rewraps its argument after recursively applying !desugarSqr!.  The inferred type
for !desugarSqr! is as follows
\begin{code}
(f :-: Square = g, Product :<: g, Functor g) =>
Fix f -> Fix g
\end{code}
Note that this captures both the action of the desugaring step (the removal of the !Square! case)
and its requirement (the presence of the !Product! case) without otherwise constraining the input or
output types.

%
%

\subsection{Further Generalization}\label{sec:further}

We conclude this section by discussing a possible further generalization of our injection and
branching operators.  As defined, our inclusion relation !f :<: g! holds only if !f! appears
somewhere in !g!.  However, if !f! is itself a coproduct, this definition may be less expressive
than we would want.  For example, we cannot show that !(A :+: C) :<: (A :+: (B :+: C))!, or even
that !(A :+: B) :<: (B :+: A)!.  The subtraction relation is similarly constrained; for example,
there is no type $\tau$ such that
\[
  \mathtt{((A \oplus B) \oplus C)} \ominus \mathtt{(A \oplus C)} = \tau.
\]
We will show how we can extend our existing definitions to account for these cases as well.

We begin with the injection function.  In this case, the intended behavior is straightforward: when
injecting a value of !(f :+: g) e!, rather than injecting the value directly, we attempt to inject
each case separately.  That is, we would add the following clause to our existing definition
(\figref{inj}), after line 9:
\begin{code}
else (f :+: g) :<: h if f :<: h, g :<: h where
  inj = inj .?. inj
\end{code}
We use the primitive branching operator to define the separate behavior for values of type !f e! and
!g e!; in each case, we rely on a recursive invocation of the injection function.

We next consider the branching combinator.  From a typing perspective, this case is appealingly
direct: we implement !f :-: (g :+: h)! as !(f :-: g) :-: h!, which looks very much like a
distributive law.  To implement this case, we would add the following clause to our existing
definition (\figref{branching}) after line 7:
\begin{code}
else f :-: (g :+: h) = (f :-: g) :-: h where
  m ? n = (m . Inl) ? ((m . Inr) ? n)
\end{code}
In implementing the branching combinator, we have three possibilities: the argument is of type !g!
(the left case handled by !m!), or it is of type !h! (the right case handled by !m!) or it is of
type !(f :-: g) :-: h! (the case handled by !n!).  We implement the branching combinator by
combining these three options.

Unfortunately, while these extensions are relatively easy to implement, they are less useful in
practice.  In particular, as we will discuss further in the following section, using the extension
to !(:-:)! always requires the introduction of explicit type signatures to avoid ambiguity in the
resulting types.

\section{The Coherence Problem}\label{sec:coherence}

In the previous section, we showed terms that constructed extensible variants (such as the terms !x!
and !y! of our simple arithmetic languages), terms that consumed extensible variants (such as the
evaluation functions !eval1! and !eval2!), and even terms that did both (such as the desugaring
function !desugarSqr!).  We have shown that the inferred types for each of these terms are suitably
general, neither constraining the order of types in coproducts nor preventing their use at larger
types.  One might be tempted to conclude that we had a complete encoding of extensible variants.

However, when we attempt to use these terms in composition, we discover an insidious problem.
Consider the innocuous term
\begin{code}
x' = eval1 x
\end{code}
We might hope that !x'! would have type !Int! and value !3!.  Trying this example, however, leads to
quite a different conclusion: that typing leaves an ambiguous type variable (say !f!), subject to
the constraints that !Sum :<: f!, !Const :<: f!, and !f :-: Const = Sum!.  In fact, we have already
observed that there are two such types (!Const :+: Sum! and !Sum :+: Const!), as these give the
distinct types !E1! and !E1'!.

This problem is pervasive.  It arises at any composition of the introduction and elimination forms
for extensible variants, that is, at any expression equivalent to $(M \? M') \, (\mathtt{inj}\, N)$
for arbitrary subterms $M,M',N$.  This difficulty also arises in the prior work on encoding
extensible variants.  It is also not immediately resolvable without losing significant
expressiveness.  For example, we might hope to add an additional functional dependency to the
$\ominus$ class fixing the order of cases:
\begin{code}
class f :-: g = h | g h -> f where ...
\end{code}
This would resolve the ambiguity, but at the cost of limiting the expressiveness of the !(?)!
combinator.  For example, we would end up with a system in which the terms $M \? N$ and $N \? M$ had
distinct and incomparable types.  We are not aware of any systems of row type (or indeed algebraic
data types in general) where the order of branches is a case expression restricts its typing.

A similar problem arises in attempts to use the extended branching operator~\secref{further}.  For
example, we might hope that it would allow us to use the following definition
\begin{code}
eval2' = cases ((evalConst ? evalSum) ?
                evalProduct)
\end{code}
However, for the added instance to apply we must conclude that the subterm !evalConst ? evalSum! has
type $(\tau \oplus \tau') \upsilon \to \mathtt{Int}$ for some particular $\tau$ and $\tau'$, but
that term can apply to arguments of types constructed from !Const :+: Sum! or !Sum :+: Const!,
leaving the type of the entire term ambiguous.

We could observe that the choice of !Const :+: Sum! or $\mathtt{Sum} \oplus \mathtt{Const}$ is
irrelevant to the result of the computation.  That is, both the terms !eval1 (x :: E1)! and
!eval1 (x :: E1')! evaluate to the same result (!3!).  On this basis, we might hope to argue that
the type checker ought to be free to make either choice without restricting the behavior of the
resulting programs or introducing incoherence, just as the type checker is free to choose the list
element type in the expression !null []!.  Unfortunately, this is not true either.  Consider the
following only-somewhat-contrived example:
\begin{code}
lefty (In (Inl _)) = True
lefty (In (Inr _)) = False
x' = (\y -> (eval1 y, lefty y)) x
\end{code}
As before, the type of !y! is ambiguous.  Suppose we left the type checker free to pick an
instantiation (we defer, for now, the question of how the type checker might make such a selection).
If it picked !E1!, !y! would be of the form !In (Inr ...)!, and !x'! would be \scantokens{!(3,
  False)!}\unskip; on the other hand, if it picked !E1'!, !y! would be of the form !In (Inl ...)!, and !x'!
would be !(3, True)!.  Thus, !lefty! is sufficient to witness the incoherence introduced by the type
checker's choice of type.

We might still hope to salvage a usable system.  We can observe that !lefty! is different from the
other eliminators we have presented: it branches on the structure of the coproduct directly, rather
than using the general branching combinator.  Terms defined using the general branching combinator,
in contrast, cannot observe whether the type checker chose !E1! or !E1'!.  Thus, by treating !(:+:)!
as an abstract type, accessible only through the !inj! and !(?)! functions, we could allow the
compiler to choose the instantiation of coproducts without compromising coherence.  Of course, the
Haskell module system is already sufficient to hide the constructors of !(:+:)!.  The only remaining
problems is how to tell the type checker which ambiguous type variables it is free to instantiate,
and how to instantiate them.

A similar problem arises in the use of Haskell numeric types.  Consider the definition
\begin{code}
z = show 1
\end{code}
We might hope to conclude that !z! has type !String! and value !"1"!.  However, a strict
interpretation of qualified types would suggest that this type was ambiguous: the constant !1! has
type !Num a => a!, and !a! is not fixed by !Show!.  Haskell includes a defaulting mechanism,
allowing this expression to type despite the ambiguity.  The defaulting mechanism defined in the
Haskell report~\cite{Haskell} is restricted to numeric classes.  We propose generalizing it to apply
to user-defined classes as well.  Consider the default defaulting declaration
\begin{code}
default (Integer, Double)
\end{code}
To generalize such declarations, we must begin by adding information about which constraints induce
default instantiations.  For example, we could make the above declaration more explicit by writing
something like
\begin{code}
default (Num Integer, Num Double)
\end{code}
This clarifies that constraints of the form !Num t!, where type variables !t! is ambiguous, should
induce defaulting.  Generalizing this idea to the multi-parameter case, we can use a similar
declaration to resolve the ambiguity present in our examples:
\begin{code}
default ((g :+: h) :-: g = h)
\end{code}
This declaration indicates that constraints of the form $f \ominus \tau = \upsilon$, where type
variable $f$ is ambiguous, should induce defaulting, instantiating variable $f$ to the type
$\tau \oplus \upsilon$.  It is easy to verify that this rule is sufficient to resolve the ambiguity
present in our examples.

Implementing an extension like this one would require consideration of a number of additional
details; we list a few of them here.  First, the type checker must confirm that defaulting
assertions are sensible at all (that is, that the instantiations do not introduce new type errors).
Second, defaulting declarations are currently limited to the module in which they occur; for our
generalized defaulting declarations to be useful, they must hold in importing modules as well.
Third, we may encounter conflicting default declarations; these should presumably generate errors at
compile time.  Most significantly, we would expect that programmers would only introduce default
declarations in cases where they did not introduce incoherence.  We cannot expect compilers to
verify such a condition automatically---but this is no different from the hope that !Eq!  instances
leave !(==)! being an equivalence relation or that !Monad! instances obey the monad laws.  Of
course, the existing defaulting mechanism is hopelessly incoherent; for example \scantokens{!show (1
  :: Integer)!} and !show (1 :: Double)! produce different output.  We can, perhaps, hope to do
better going forward.

\section{Extensible Variants with Type Families}\label{sec:tf}

In the previous sections, we have developed a coproduct-based implementation of extensible variants,
including both overloaded injection and branching operators, and have compared it to other
approaches to extensible variants.  However, our implementations rely on instance chains, an
extension of the Haskell class system only available in prototype implementations.  In this section,
we translate our implementations to use closed type families~\cite{Eisenberg14}, a related extension
of the GHC type system.  Unlike instance chains, closed type families operate purely at the type
level---they do not directly determine method implementations.  Thus each instance chain in the
original implementations will correspond to (at least) two components in the translation: first, a
closed type family which searches the possible solutions arising from the instance chain and (if
successful) produces a type-level witness that the predicate holds; and, second, a (standard)
Haskell type class which uses the type-level witness constructed by the closed type family to
determine method implementations.

\subsection{Injection}

\begin{figure}
\begin{codef}
data Yep; data Nope

type family IsIn f g where
  IsIn f f       = Yep
  IsIn f (g :+: h) = Or (IsIn f g) (IsIn f h)
  IsIn f g       = Yep

type family Or b c where
  Or Nope Nope = Nope
  Or b c       = Yep
\end{codef}
\caption{Types not in variants.}\label{fig:notin-tf}
\end{figure}

We begin by considering the !In! class (\figref{in}).  Unlike the other classes we will consider,
!In! does not provide any methods.  This simplifies the translation, as we can rely entirely on type
families.  The translation of !In! is given in \figref{notin-tf}.  We originally defined !In! as a
relation on types, relying on negative predicates to describe types not in the relation.  We
translate !In! as its characteristic function !IsIn!: !IsIn f g! rewrites to !Yep! if !f! is a
summand of !g!, and to !Nope! if it is not.  The second and third instances of !In! describe a
disjunction; we implement this with a new type family !Or!, which rewrites to !Nope! if both of its
arguments do, and to !Yep! otherwise.

\begin{figure}
\begin{codef}
data Refl; data L x; data R x

type family Into f g where
  Into f f       = Refl
  Into f (g :+: h) = Ifi (Into f g) (IsIn f h)
                       (Into f h) (IsIn f g)
  Into f g       = Nope

type family Ifi lp inr rp inl where
  Ifi Nope inr Nope inl = Nope
  Ifi Nope inr rp Nope  = R rp
  Ifi lp Nope rp inl    = L lp
  Ifi lp inr rp inl     = Nope
\end{codef}
\caption{Finding types in sums.}\label{fig:in-tf}
\end{figure}

The !(:<:)! class (\figref{inj}) defines the !inj! method.  In translating !(:<:)! to type families,
we will need to define both a type family (which implements instance chain proof search, computing a
type-level witness of an !(:<:)! proof) and a type class (which builds the implementation of the
!inj! method from the type-level witness). \figref{in-tf} gives the type family.  We begin by
introducing type-level witnesses of !(:<:)! proofs.  !Refl! denotes a proof of !In f f!; it
corresponds to the first clause in the !(:<:)! definition.  !L p! denotes a proof of \scantokens{!In
  f (g :+: h)!} if !f! is found in !g!, where !p! is the witness that !f! is a summand of !h!.  This
corresponds to the second clause in the !(:<:)!  definition.  Note that in our translation, we only
need to track those constraints that contribute to the implementation of !inj!, so we do not include
the results of !IsIn! in our witnesses.  !R! is similar, but for the case where !f! is found in !h!.
For example, we expect that !In B ((A :+: B) :+: C)! would rewrite to !L (R Refl)!.  Finally, we
reuse !Nope! to denote the proof that !In f g! cannot hold; for example, !In D (A :+: B)! should
rewrite to !Nope!.

The type family !Into f g! implements !f :<: g! proof search.  The first and third equations (lines
4 and 7) are straightforward, handling the reflexive case and the case where argument !g! is not a
sum.  The second equation (lines 5--6) must handle all the cases where !g! is a sum !gl :+: gr!;
these correspond to the second, third, and some uses of the fourth clause in the !(:<:)!
definition.  The branching is delegated to an auxiliary class !Ifi lp inr rp inl! where !lp!
(respectively !rp!)  witnesses a proof that !f! can be injected into !gl! (respectively !gr!) while
!inr!  (respectively !inl!)  witnesses a proof that !f! appears (possibly more than once) in !gr!
(respectively !gl!).  The second and third equations (lines 11--12) are the successful cases, in
which !f! appears on one side of the sum but not the other.  The first equation (line 10) handles
the case where !f! appears on neither side of the sum, while the last (line 14) handles the case
where it appears on both.  For example, we can see that !In A (A :+: B)!  would rewrite to
\scantokens{!Ifi Refl Yep Nope Nope!}\unskip, which would write to !L Refl!, while !In A (A :+: A)!
rewrites to !Ifi Refl Yep Refl Yep!, which rewrites to !Nope!.

\begin{figure}
\begin{codef}
class Inj f g p where
  injp :: p -> f e -> g e

instance Inj f f Refl where
  injp _ = id

instance Inj f g p => Inj f (g :+: h) (L p) where
  injp (_ :: L p) = Inl . injp (undefined :: p)

instance Inj f h p => Inj f (g :+: h) (R p) where
  injp (_ :: R p)  = Inr . injp (undefined :: p)

inj :: forall f g e.
       (Inj f g (Into f g)) => f e -> g e
inj = injp (undefined :: Into f g)
\end{codef}
\caption{Overloaded injection function.}\label{fig:inj-tf}
\end{figure}

We can use the results of !Into! to define the injector !inj!, as shown in \figref{inj-tf}.  We
begin by defining a class !Inj f g p! (lines 1--2); !p! is the witness of the proof that !f! is a
summand of !g!.  This class has a single method !injp!; in addition to an argument of type !f e!, it
takes an argument of type !p!.  The are three instances of this class, corresponding to the three
constructors of inclusion proofs.  The case for !Refl! (lines 4--5) is straightforward.  The cases
for !L p! (lines 7--8) and !R p! (lines 10--11) are similar; we will describe the first.  If the
witness is of the form !L p!, then the injector should inject into the left-hand component of the
coproduct.  The instance can thus assume that the second argument is a coproduct !g :+: h!, and
assumes that !Inj f g p! holds (effectively assuming that !p! withesses that !f! is a summand of
!g!).  We define !injp! as the composition of !Inl! and the injector of !f! into !g! given by !p!.
Referring to the latter requires a value of type !p!; as these values are used solely for carrying
types, !undefined!  will do.  Finally, we can define a function !inj! that hides the type-level
witnesses (lines 13--15); again, we can use !undefined! as a value of type !Into f g!.  Note that we
rely on GHC's scoped type variables extension to allow us to refer to !p!  (in the !Inj! instances)
and !f! and !g!  (in the definition of !inj!).

The definition of !Into! and !Inj! contain overlapping structure, such as the assumption
!Inj f g p! in the instance of !Inj! for witnesses !L p!.  Suppose that there were a bug in the
implementation of !Into! such that !Into A (A :+: B)! rewrote to !L Refl! instead of !R Refl!.  The
definitions would still be accepted by GHC; however, in attempting to use !inj! at type
!A e -> (A :+: B) e!, the typechecker would have to discharge an instance !Inj A B Refl!.  There is
no instance to do so, leaving an (unsolvable) constraint in the resulting type.  This demonstrates
that, even if the !Into! and !Inj! classes do not align, type safety is not compromised.  On the
other hand, !Into! also assures invariants that are not necessary for type safety---for example, it
rules out arbitrary injections like !In A (A :+: A)!.  Bugs in these invariants would not introduce
problems in the interplay between !Into! and !Inj!.  For example, suppose that !Into A (A :+: A)!
rewrote to !R Refl! (instead of !Nope!).  We would then be able to use !inj! at type
!A e -> (A :+: A) e!; it would correspond to !Inr!.

\subsection{Branching}

We next translate the !(:-:)! class, our implementation of branching (\figref{branching}).  This
translation follows broadly the same pattern as the translation of !(:<:)!: we introduce a type
family !Minus! that implements the search for !(:-:)! proofs, and a type class !Without! that
implements the branching combinator !(?)! based on the witnesses produced by !Minus!.  However, there
is one significant new complication.  The !(:-:)! class has a functional dependency: if the
predicate $\tau \ominus \tau' = \upsilon$ holds, the combination of $\tau$ and $\tau'$ determine
$\upsilon$.  Correspondingly, our translation of !(:-:)! will compute not just a proof, but also the
determined type $\upsilon$.

\begin{figure}
\begin{codef}
data Onl (h :: * -> *)
data Onr (h :: * -> *)
data Le (g :: * -> *) p
data Ri (f :: * -> *) p

type family Minus f g where
  Minus f f       = Nope
  Minus (f :+: g) f = Onl g
  Minus (f :+: g) g = Onr f
  Minus (f :+: g) h = Ifm g (Minus f h) (IsIn f g)
                        f (Minus g h) (IsIn f h)
  Minus f g       = Found f

type family Ifm g lp inr f rp inl where
  Ifm g Nope inr f Nope inl = Nope
  Ifm g Nope inr f rp Nop e = Onr f rp
  Ifm g lp Nope f rp inl    = Onl g lp

type family OutOf p where
  OutOf (Onl x)  = x
  OutOf (Onr x)  = x
  OutOf (Le f p) = OutOf p :+: f
  OutOf (Ri f p) = f :+: OutOf p
\end{codef}
\caption{Subtracting types from sums.}\label{fig:minus-tf}
\end{figure}

The type-level translation of !(:-:)! is given in \figref{minus-tf}.  As in the last section, we
begin with type-level witnesses of proofs of $\tau \ominus \tau' = \upsilon$.  !Onl h! and !Onr h!
witness the base cases, where !h! captures the remaining type.  For example, we would expect
!Minus (A :+: B) A! to rewrite to !Onl B!; the evidence constructors are named by the location of
the subtrahend, not the location of the remainder.  !Le g p! witnesses a proof of
!(f :+: g) :-: h = k!  where !h! is found in !f!; the witness includes both !g!, one component of
the result type !k!, and the witness !p! that !h! can be subtracted from !f!. !Re f p! is similar,
but accounts for the case when !h! is found in !g!.  For example, we would expect
!Minus ((A :+: B) :+: C) B! to rewrite to !Le C (Onr A)!.  Note that as the results of !IsIn! do not
contribute to the implementation of !(?)!, we have omitted them from the witnesses of !Minus!.  The
implementation of !Minus! is mostly unsurprising.  Lines 7--9 and 12 contain base cases.  The
recursive cases are all captured in lines 10--11, and deferred to the auxiliary class !Ifm!.  In the
type !Ifm g pf ing f pg inf!, !g! and !f! are the summands of the original coproduct, !pf! and !pg!
the result of subtracting !h! from !f! and !g!, respectively, and !ing! and !inf! capture whether
!h! appears at all in !g! and !f!.  The first equations (line 15) captures the case where !h!
appears on neither side of the sum; the remaining equations capture the cases where !h! appears on
one side but not the other.

The typing of !(?)! depends upon the result type !h!; to express it, we define an additional type
function !OutOf! to extract the computed result type from a !Minus! witness (lines 19--23).  For
example, given that !Minus ((A :+: B) :+: C) B! rewrites to !Le C (Onr A)!, we expect
\scantokens{!OutOf (Le C (Onr A))!} to rewrite to !A :+: C!, the components of the original
coproduct remaining after removing !B!.  Note that the distinction between !Onl! and !Onr! is not
significant in defining the result type; we need only distinguish the cases to assure that the
definition of !(?)! is unambiguous.

\begin{figure}
\begin{codef}
class Without f g p where
  (??) :: (g e -> r) -> (OutOf p e -> r) -> p
       -> f e -> r

instance Without (f :+: g) f (Onl g) where
  (m ?? n) _ = m .?. n

instance Without (f :+: g) g (Onr f) where
  (m ?? n) _ = n .?. m

instance Without f h p =>
         Without (f :+: g) h (Le g p) where
  (m ?? n) (_ :: Le g p) =
     (m ?? (n . Inl)) (undefined :: p) .?. (n . Inr)

instance Without g h p =>
         Without (f :+: g) h (Ri f p) where
  (m ?? n) (_ :: Ri f p) =
     (n . Inl) .?. (m ?? (n . Inr)) (undefined :: p)

(?) :: forall f g e r. Without f g (Minus f g)
    => (g e -> r) -> (OutOf (Minus f g) e -> r)
    -> f e -> r
m ? n = (m ?? n) (undefined :: Minus f g)
\end{codef}
\caption{Overloaded branching combinator.}\label{fig:branch-tf}
\end{figure}

We can now implement the branching combinator itself, shown in \figref{branch-tf}.  As for !Inj!,
the !Without! class (lines 1--3) has not only !f! and !g! parameters, but also a witness !p! of
!Minus f g!.  However, !p! now appears twice in the type of !(??)!: not just to direct the !Without!
class, but also to give the type of the right-hand argument of !(??)!.  The base cases (lines 5--9)
are straightforward; note that the instances do not apply given the wrong base-case witness.  The
recursive cases are more interesting.  We consider the case for !Le g p! (lines 11--14); the !Ri!
case (lines 16--19) is similar.  We begin with the types of !m! and !n!.  Suppose that
\scantokens{!Minus (f :+: g) h = Le g p!}\unskip.  We know that !m :: h e -> r! and \scantokens{!n :: OutOf
  (Le g p) e -> r!}\unskip; from the definition of !OutOf!, we know that !OutOf (Le g p) = f' :+: g! where
we assume that !OutOf (Minus f h) = f'!.  Finally, we consider the possible arguments to
\scantokens{!(m ?? n) p!}\unskip.  If the argument is !Inr x!, then !x :: g e!; this is the right-hand case
handled by !n!.  On the other hand, if the argument is !Inl x!, then we know it is handled by either
!m! or by the left-hand case of !n!, and we rely on a recursive call to !(??)! to determine which
case applies.  As in the definition of !Inj!, we rely on a parameter tracking the witness to
disambiguate the recursive call.

Finally, we can define a wrapper function !(?)! which hides the need for a !Minus! witness.  The
definitions of !Minus! and !Without! are intertwined: !Without! relies on !Minus! witnesses being
correctly constructed, and assumes (without proof) that !Minus! properly enforces its invariants.
For example, suppose that a bug in the definition of !Minus! resulted in !Minus (A :+: A)! rewriting
to !Onr A!.  We could then have
\begin{code}
(?) :: (A e -> r) -> (A e -> r) -> (A :+: A) e -> r
\end{code}
where in !m ? n!, !m! is applied to !Inr! cases and !n! to !Inl! cases.

\subsection{Discussion}

We conclude by comparing our translations of $\olessthan$ and $\ominus$ with the original, and
discussing some issues arising from the translation.

For ground types (i.e., types without type variables), the translations are equally expressive.
That is, for any ground types $\tau,\tau',\upsilon$, we can prove $\tau\olessthan\upsilon$ if and
only if we can prove $\predht{Inj}{\tau\,\upsilon\,(\predht{In}{\tau\,\upsilon})}$, and $\tau
\ominus \tau' = \upsilon$ if and only if we can prove $\predht{Without}{\tau \, \tau' \,
  (\predht{Minus}{\tau \, \tau'})}$ such that $\predht{OutOf}{(\predht{Minus}{\tau\,\tau'})} \sim
\upsilon$.  The correspondence is not as close in the presence of type variables.  For example,
using the instance chains encoding allows the following type for !inj!:
\begin{code}
inj :: In f g fails => f e -> (f :+: g) e
\end{code}
In this case, the !In f g fails! assumption is sufficient to discharge the constraint \scantokens{!f
  :<: (f :+: g)!}\unskip.  However, the same does not hold for the implementation using type families.
That is, we cannot show the typing
\begin{code}
inj :: IsIn f g ~ Nope => f e -> (f :+: g) e
\end{code}
Matching in closed type families is based on infinitary unification (even though GHC does not permit
infinite types); thus, the type !In f (f :+: g)! can rewrite either to !L Refl! (relying on the
assumption that !IsIn f g ~ Nope!) or to !Refl! (relying on the unification $\mathtt{f} \sim
\mathtt{f} \oplus \mathtt{g}$).  Because of this ambiguity, the !Into! type function does not
rewrite until !f! and !g! have concrete instantiations.  Thus, we can conclude that our translation
in terms of type families is not quite as expressive as the original.  However, it is unclear how
significant this loss of expressiveness would be in practice.  While it results in more complex
types for polymorphic functions, we have not found any programs which can type under one scheme but
are do not type (with any type scheme) under the other.

The !(:-:)! class has two functional dependencies: in a predicate !(:-:) f g h!, !f! and !g! are
sufficient to determine !h!, and !f! and !h! are sufficient to determine !g!.  In our encoding, we
have only made use of the first functional dependency.  However, we could add the second to the
definition of !(:-:)! (\figref{branching}) without requiring any other changes; the existing
instances satisfy that dependency as well.  The case is not as clear for !Minus! and !Without!,
however.  It is true that, if $\mathtt{OutOf}\,(\mathtt{Minus}\,\tau\,\tau') = \upsilon$, then
$\mathtt{OutOf}\,(\mathtt{Minus}\,\tau\,\upsilon) = \tau'$.  However, in defining !Without! (and
thus !(?)!), we have chosen which parameter to be determined: the generation of the witness !p! and
computation of !h! go hand-in-hand.  We could certainly define a version of !Without! in which the
other parameter were determined, but this would have to be a separate definition, resulting in a
different branching combinator.


\section{Related Work}\label{sec:related}

Blume et al.~\cite{Blume06} give an ML-like language extended with polymorphic records and variants.
Their system allows individual cases to be defined independently and combined (as with our !(?)!
operator); however, their type system distinguishes first-class cases from functions and introduces
a distinct elimination form for them.  They exploit the duality of products and coproducts to
compile extensible variants into extensible records, and then into efficient index-passing code.
Garrigue~\cite{Garrigue08} gives a system of polymorphic variants, implemented in Ocaml.  His system
does not support extensible variants directly.  However, Blume et al. observe that, by modifying his
type system somewhat, his compilation techniques could be adapted to support extensible variants.

Row typing was originally introduced by Wand~\cite{Wand87}, as a mechanism for typing extensible
records (and thus, objects with inheritance).  His system did not include any way to restrict the
labels that appeared in a given row; this resulted in an incompleteness in his type inference
algorithm.  R\'emy~\cite{Remy89,Remy92} proposed a modification of Wand's system that incorporated
presence information into rows, and so could express the absence of a label.  R\'emy's system thus
repairs the incompleteness in Wand's type inference algorithm.  Gaster and Jones~\cite{Gaster96}
give a version of row typing that makes use of predicates to exclude types from rows, rather than
incorporating absence information into the rows directly.  This simplifies the form of types.  They
show that their system also enjoys complete type inference.

There is a large and varied literature on using type classes to encode extensible records and
variants.  Liang et al.~\cite{Liang95} is the earliest we are aware of; their approach requires
hardwiring recursive uses of data types, but otherwise supports overloaded injection and projection
operators.  Kiselyov et al.~\cite{Kiselyov04} focus on heterogeneously-typed lists, and define
type-directed lookup and removal operators.  Their lists can be viewed as extensible records, and
the type signatures of their operators parallel ours (albeit limited to list-like structures).  They
show how their approach can be adapted to work with labeled types, but do not address variants
directly.  Swierstra~\cite{Swierstra08} generalized the approach of Liang et al. to support
recursive types without hardwiring, but relies on introducing new type classes for each function
consumes extensible variants.

Bahr~\cite{Bahr14} describes an approach to extensible variants implemented using closed type
families.  His approach is initially similar to our type-family-based approach.  However, there are
several key differences.  He defines a projection operator similarly to that of Liang et al., rather
than defining a branching combinator as we do.  Defining the projection operator in terms of
branching is direct:
\begin{code}
prj = Just ? const Nothing
\end{code}
Defining branching in terms of projection is not as straightforward.  Bahr accomplishes it by
generalizing injection to deconstruct coproducts, similar to the further generalizations of
injection we discussed~\secref{further}.  He can then use his injector to rearrange coproducts and
standard \texttt{case} statements for branching.  For example, given a term !x! of type $(f \oplus
(g \oplus h)) \, e$ and a branch !m! of type $g \, e \to r$, he can use !inj! to get a term of type
$(g \oplus (f \oplus h)) \, e$, and then do case analysis on that term, applying !m! in the !Inl!
branch.  Our approach differs in two important ways.  First, Bahr's approach sometimes leaves
ambiguities that are not present in our approach; we do not know if they would be resolved by a
similar defaulting mechanism to the one we have proposed.  Second, his approach relies on leaving
the implementation of the coproduct type exposed, whereas we can treat !(:+:)! as an abstract type.


Morris and Jones~\cite{MorrisJones10} observed that instance chains could be used to define a more
expressive coproduct injector, but do not completely rule out ambiguous coproducts.
Morris~\cite{Morris13} gives a version that rules out ambiguous coproducts, and gives a version of
the branching combinator.  That work does not consider the coherence problems, and does not translate
their implementation into closed type families.

\section{Conclusions}\label{sec:conclusions}

We have described a new encoding of extensible variants in Haskell, based on overloaded injection
and branching operators, and have given two implementations of our encoding, one using instance
chains and one using closed type families.  We have compared the expressiveness of our system to
those based on row types, identified a source of ambiguity in our (and others')  encodings but not
in row type systems, and have proposed a generalized defaulting mechanism to resolve this
ambiguity.  We conclude by discussing future directions for language design and research.

We have focused exclusively on extensible variants in this paper.  We believe our approach would be
equally applicable in a number of other contexts.  Most obviously, we could apply them to build
extensible records, but we also imagine they would have applicability in encoding effect type
systems.  In particular, we think there may be overlap between our typing of
desugaring~\secref{branch} and the typing of effect handlers.

In implementing !inj! and !(?)! using closed type families, we were able to translate our instance
chain-based implementations fairly directly.  This raises the question about whether a translation
from instance chains to closed type families can be defined in general, and whether it could be
automated.  Such a translation would greatly reduce the cost of providing instances chains in GHC.
Even if not all instance chains could be translated, we believe it would still contribute to
identify those instance chains which could be translated, and provide automated translation in those
cases.

The broader question raised by this work is how best to provide features like extensible variants in
Haskell.  We believe that there are three possible answer to this question.
\begin{itemize}
\item We may conclude that all of the approaches to encoding extensible variants are simply too
  complex, relying on numerous extensions of existing type and class systems, and are unlikely to be
  useful in practice.  By comparison, row typing is well studied, has been implemented in Haskell
  systems in the past, and may require less overall complexity (even if it does necessarily touch
  the core type system).
\item Alternatively, we may conclude that the status quo is fine.  While the encodings are complex,
  this is to be expected for complex features.  The present work demonstrates that the encodings can
  be sufficiently expressive, based only on existing features.  Finally, while the introduced
  ambiguities are unpleasant, we may claim that programmers ought to be writing type signatures
  anyway.
\item Finally, we may conclude that we are most of the way there, and that small additions, such as
  our generalized defaulting mechanism, should get us the rest of the way.  Features like instance
  chains or closed type families are generally useful, not simply for encoding extensible records
  and variants.  Further, we suspect that ambiguity issues like the ones we encounter will appear
  in other contexts as well.  Solutions to these problems will enable more features than just
  extensible variants.
\end{itemize}
Unsurprisingly, perhaps, we take the third perspective.  We acknowledge that it is at least partly a
matter of taste.  We hope that further development of these ideas, including investigation of the
causes of ambiguity and techniques for assuring coherence, can shed further light on these options
and lead to more modular Haskell programs in the future.

\section*{Acknowledgments}

This work began with the advice of Mark Jones.  Thanks to Ben Gaster
and to the anonymous referees for feedback on earlier versions of this paper.  Morris was supported
by EPSRC grant EP/K034413/1.

\bibliographystyle{abbrvnat}
\bibliography{cites}{}
\end{document}

\appendix

\section{Instance Chains, Formally}
\newcommand{\prog}[2]{#1|#2}
\newcommand{\progprec}{\leq}

\todo{
\begin{itemize}
\item Include type system?  (Problem: OML doesn't include type annotations, but need type
  annotations to disambiguate.)
\item Syntax
\item Unmatched-positive rule (needs to take fundeps into account)
\end{itemize}}

\begin{figure*}
\begin{gather*}
\infbox{\irule[\erule{Atom}]
              {Q \subseteq P};
              {\prog A X \vdash P \entails Q}}
\isp
\infbox{\irule[\erule{Conj}]
              {\bigwedge\Set{\prog A X \vdash P \entails \pi \mid \pi \in Q}};
              {\prog A X \vdash P \entails Q}}
\isp
\infbox{\irule[\erule{Inst}]
              {\alpha \in A}
              {\alpha \vdash P \entails \pi};
              {\prog A X \vdash P \entails \pi}}
\\[3px]
\infbox{\irule[\erule{Prove}]
              {(Q \then \pi) \in \gr{\clause{}{\seq t}{Q'}{\pi'}}_\pi}
              {\prog A X \vdash P \entails Q};
              {(\chain{}{\seq t}{Q'}{\pi'}{\alpha}) \vdash P \entails \pi}}
\isp
\infbox{\irule[\erule{Contra}]
              {(Q \then \pi'') \in \gr{\clause{}{\seq t}{Q'}{\pi'}}_\pi}
              {\prog A X \vdash P \entails \overline{Q}}
              {\alpha \vdash P \entails \pi};
              {\chain{}{\seq t}{Q'}{\pi'}{\alpha} \vdash P \entails \pi}}
\\[3pt]
\infbox{\irule[\erule{Unmatched}]
              {\gr{\clause{}{\seq t}{Q'}{\pi'}}_\pi = \emptyset}
              {\alpha \vdash P \entails \pi};
              {(\chain{}{\seq t}{Q'}{\pi'}{\alpha}) \vdash P \entails \pi}}
\end{gather*}
\caption{Entailment rules for instance chains}
\end{figure*}

\end{document}